\documentclass[aps,prl,twocolumn,amsmath,amssymb,nofootinbib,superscriptaddress]{revtex4}

\newcommand{\ket}[1]{|#1\rangle}

\usepackage[pdftex]{graphicx}
\usepackage{mathrsfs}

\begin{document}

\bibliographystyle{apsrev}

\title{Optimising number resolving photo-detectors using classical post-processing}

\author{Peter P. Rohde}
\email[]{dr.rohde@gmail.com}
\homepage{http://peterrohde.wordpress.com}
\affiliation{Centre for Quantum Computation and Communication Technology, School of Mathematics and Physics, University of Queensland, QLD 4072, Australia}

\date{\today}

\frenchspacing

\begin{abstract}
Many present day quantum optics experiments, particularly in optical quantum information processing, rely on number-resolving photo-detection as a basic building block. In this paper we demonstrate that a simple classical optimisation technique can sometimes be employed to post-process the detector signature and improve the confidence of the measurement outcome in the presence of photon-number errors such as loss or dark-counts. While the regime in which this technique is applicable is rather restrictive, and will likely not be very useful for the large-scale quantum information processing applications of the future, the ideas presented might be employed in some present-day experiments where photo-detectors are typically very poor.
\end{abstract}

\maketitle

\section{Introduction}

Number-resolving photo-detection forms the basis of many quantum optics experiments, particularly in the emerging field of optical quantum information processing \cite{bib:KLM01,bib:Kok05,bib:KokLovett11,bib:NielsenChuang00}. Ideally we aim to employ detectors that can distinguish, with high fidelity, between different Fock states. However, in practise, unavoidable effects such as loss and dark-counts corrupt the photo-detection process, lowering the fidelity of our measurement results.

In this paper we present a straightforward technique, which in some parameter regimes, allows classical post-processing of the measurement signal to improve the fidelity of the measurement outcome. We present two example applications for this technique, including a simple application of practical interest -- heralded Fock state preparation using parametric down-conversion (PDC). We show that in some circumstances our technique can indeed improve the fidelity of the heralded state. Regrettably, improvement is only possible in the regime where the detector is already behaving badly (e.g. lossy or with significant dark-counts). However, given that many present-day experiments employ poor photo-detection, our technique may nonetheless be applicable.

We contrast our technique to the matrix inversion techniques that are used to reconstruct photon-number distributions \cite{bib:Lee04,bib:Achilles04}. In such techniques we take a large number of measurements to determine a vector $\vec{v}_\mathrm{meas}$ comprising probabilities of detecting different photon numbers. The relationship between the actual photon-number distribution and the measured one can be related by a matrix $M$, which captures information about the number-resolving power of the detector, including losses and dark-counts, $\vec{v}_\mathrm{meas} = M \vec{v}_\mathrm{act}$. Then, applying an inversion and fitting technique, the actual photon-number distribution can be inferred from a measured distribution. On the other hand, the technique we present here is intended for \emph{single shot} measurements, i.e. how do we optimise a single measurement rather than an ensemble of measurements? This important difference makes our technique more applicable to quantum information processing applications where a single measurement is used either for feed-forward or the measurement of the output state.

A similar technique was discussed in by O'Sullivan \emph{et al.} \cite{bib:OSullivan08}, where they focussed on the ability of lossy detectors to generate Fock states via heralded PDC. While we discuss this example also, we differentiate the present work by specifically focussing on the detector optimisation process rather than the state preparation procedure. In particular, we derive \emph{optimisation maps}, which show, for a given detector and known number distribution, how to map raw detector signatures to optimised detector signatures. In other words, the optimisation maps tell us the extent to which each measurement signature must be compensated. This technique may subsequently be used in applications broader than just heralded state preparation. We also focus on the \emph{measurement fidelities}, which tell us the confidence that our optimised signatures are correct. Finally, we demonstrate that the form of the known number distribution has a strong effect on the optimisation map and measurement fidelities, effectively biasing the regimes in which optimisation is effective. Thus, the applicability of such a technique is highly application-dependent.

\section{Detector model}

In terms of the photon-number degree of freedom (i.e. ignoring spatio-spectral effects \cite{bib:RohdeRalph06b}), the detector can be fully characterised using a matrix of conditional probabilities $P(m|n)$, the conditional probability that $m$ photons were detected given that $n$ were incident \cite{bib:Lee04,bib:RohdeWebb07}. This gives us full information about how the detector reacts to different input states.

We will focus on a model consisting of an ideal number-resolving photo-detector preceded by a loss channel and subject to dark-counts.  The loss rate is $p_\mathrm{loss}$ and the dark-count rate $p_\mathrm{DC}(i)$, where $i$ is the number of dark-counts \cite{bib:KokLovett11}. Then, we can write,
\begin{equation} \label{eq:detector_model}
P(m|n) = \sum_{d=0}^\infty p_\mathrm{DC}(d) \binom{n}{m-d}(1-p_\mathrm{loss})^{m-d}{p_\mathrm{loss}}^{n-m+d},
\end{equation}
and we assume that $p_\mathrm{DC}$ is a Poissonian distribution \cite{bib:Lee04},
\begin{equation}
p_\mathrm{DC}(d) = e^{-\lambda}\frac{\lambda ^d}{d!},
\end{equation}
where $\lambda$ is the dark-count rate and is equal to the average number of dark-counts \cite{bib:Lee04}. Thus, we are summing over all combinations where the sum of the number of dark-counts, $d$, and the number of photons minus those lost, $m-d$, is equal to the number of measured photons $m$.

\begin{figure}[!htb]
\includegraphics[width=\columnwidth]{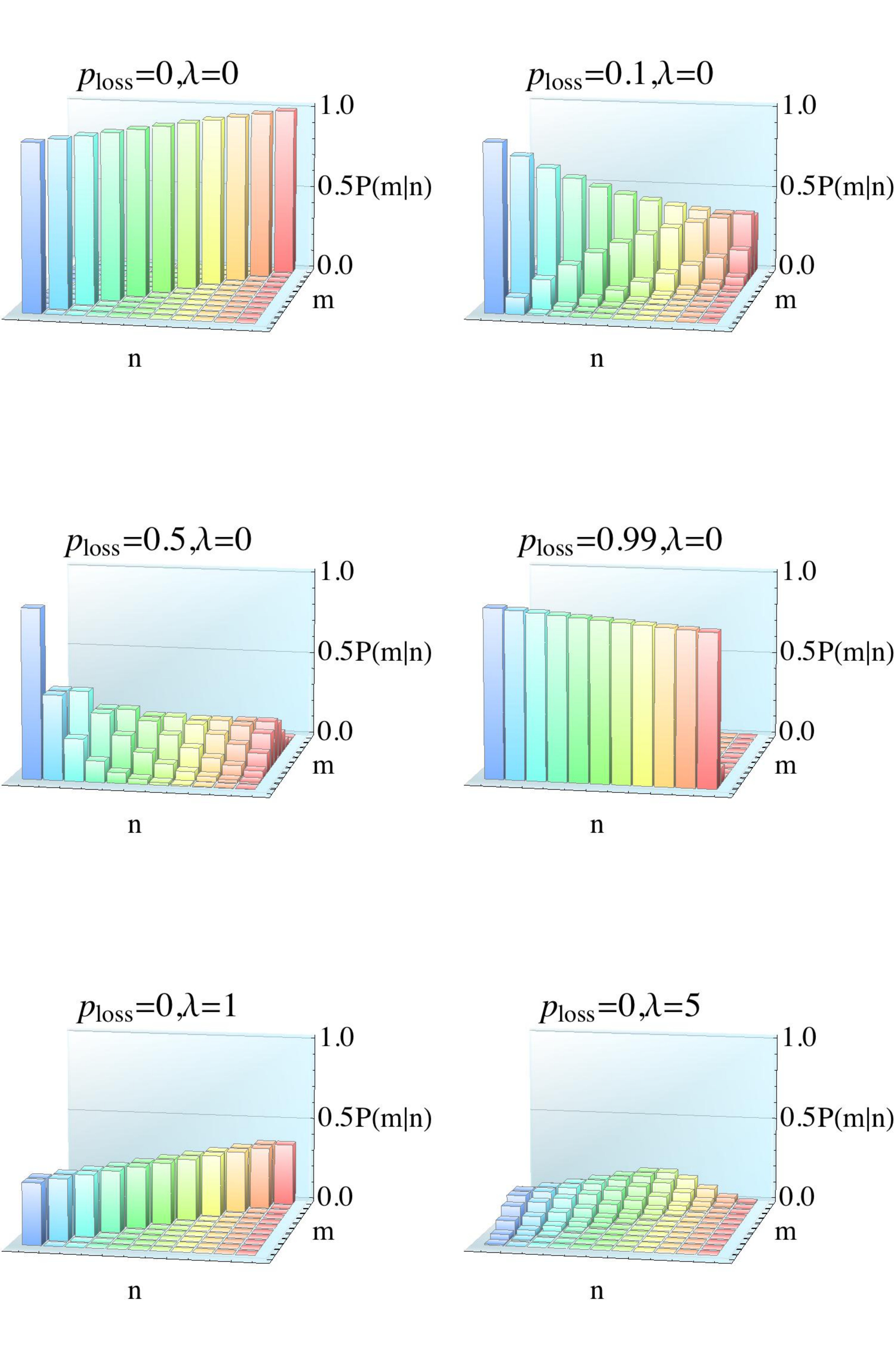}
\caption{(Colour online) $P(m|n)$ matrices for a ideal, lossy and dark-county number-resolving detectors. $m=n=0$ is front-left in each inset.} \label{fig:pmn}
\end{figure}

In Fig. \ref{fig:pmn} we show the $P(m|n)$ matrix for our detector in the presence of different loss and dark-count rates. For a perfect detector (i.e. \mbox{$p_\mathrm{loss}=p_\mathrm{DC}=0$}), the matrix is the identity matrix since every photon number input state perfectly maps to the desired signature. However, as losses and dark-counts are introduced, off-diagonal elements emerge, corresponding to incorrect measurement signatures. In the case of very high loss rates the peaks are localised around the $m=0$ axis, i.e. we always measure zero photons since everything is lost. On the other hand, in the case of high dark-count rates the peaks become very delocalised since a large number of different measurement outcomes are possible as a result of the unpredictable introduction of different numbers of dark-counts.

\section{Bayesian optimisation of detector signatures}

We have argued (following Ref. \cite{bib:RohdeWebb07}) that the operation of a number-resolving photo-detector can be fully characterized by the $P(m|n)$ matrix. However, practically we are more interested in the converse conditional probability, $P(n|m)$. That is, what is the likelihood that the incident state has some number of photons, given a particular measurement outcome. This in effect gives us the fidelity of the measurement result.

This first step can be achieved using Bayes' Law to invert the conditional probability distribution,
\begin{eqnarray}
P(n=n_0|m=m_0) &=& \frac{P(m=m_0|n=n_0)P(n=n_0)}{P(m=m_0)} \nonumber \\
&=& \frac{P(m=m_0|n=n_0)P(n=n_0)}{\sum_i P(m=m_0|n=i)P(n=i)}. \nonumber \\
\end{eqnarray}

Next we wish to optimise our signature. Specifically, for a given raw measurement signature $m$, we define the $n$ which maximises $P(n|m)$ as the optimised signature. So, for example, if our detector gives us the measurement outcome $m=2$, and then we find that $P(n|2)$ is maximised for $n=3$, then we define \mbox{$m_\mathrm{opt}=3$} as the optimised signature. Thus, by reinterpreting $m=2$ as \mbox{$m_\mathrm{opt}=3$}, we have improved the confidence of our measurement outcome. Specifically,
\begin{equation}
m_\mathrm{opt}(m) = \mathrm{argmax}_n P(n|m).
\end{equation}
This idea was first employed in Ref. \cite{bib:OSullivan08}. It can easily be seen that the state of highest confidence will always be a Fock state.

\section{Example applications}

As an example test-bed for our technique, we consider two scenarios: (1) heralded preparation of Fock states using conditioned PDC sources, which is the staple for preparing heralded Fock states in present-day single photon experiments, and (2) a uniform number distribution.

\begin{figure}[!htb]
\includegraphics[width=0.6\columnwidth]{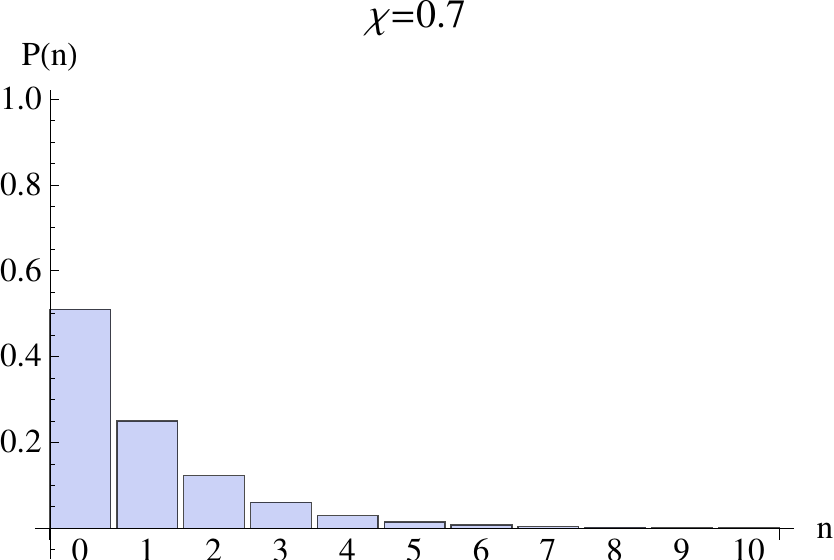}
\caption{(Colour online) Photon number distribution, $P(n)$, for a PDC state with $\chi=0.7$.} \label{fig:pn}
\end{figure}

The state emanating from a PDC can be expressed as a power series,
\begin{equation}
\ket{\psi_\mathrm{PDC}} = \sqrt{1-\chi^2} \sum_n \chi^n \ket{n}_s\ket{n}_i,
\end{equation}
where $s$ and $i$ are the signal and idler modes respectively. Thus, measuring the signal mode, the expected photon number distribution is given by
\begin{equation}
P(n) = (1-\chi^2)\chi^{2n}.
\end{equation}
The photon number distribution for $\chi=0.7$ is illustrated in Fig. \ref{fig:pn}. For the uniform distribution we assume \mbox{$P(n)=0.1$} in the range $0\dots9$, otherwise $0$.

\begin{figure}[!htb]
\includegraphics[width=\columnwidth]{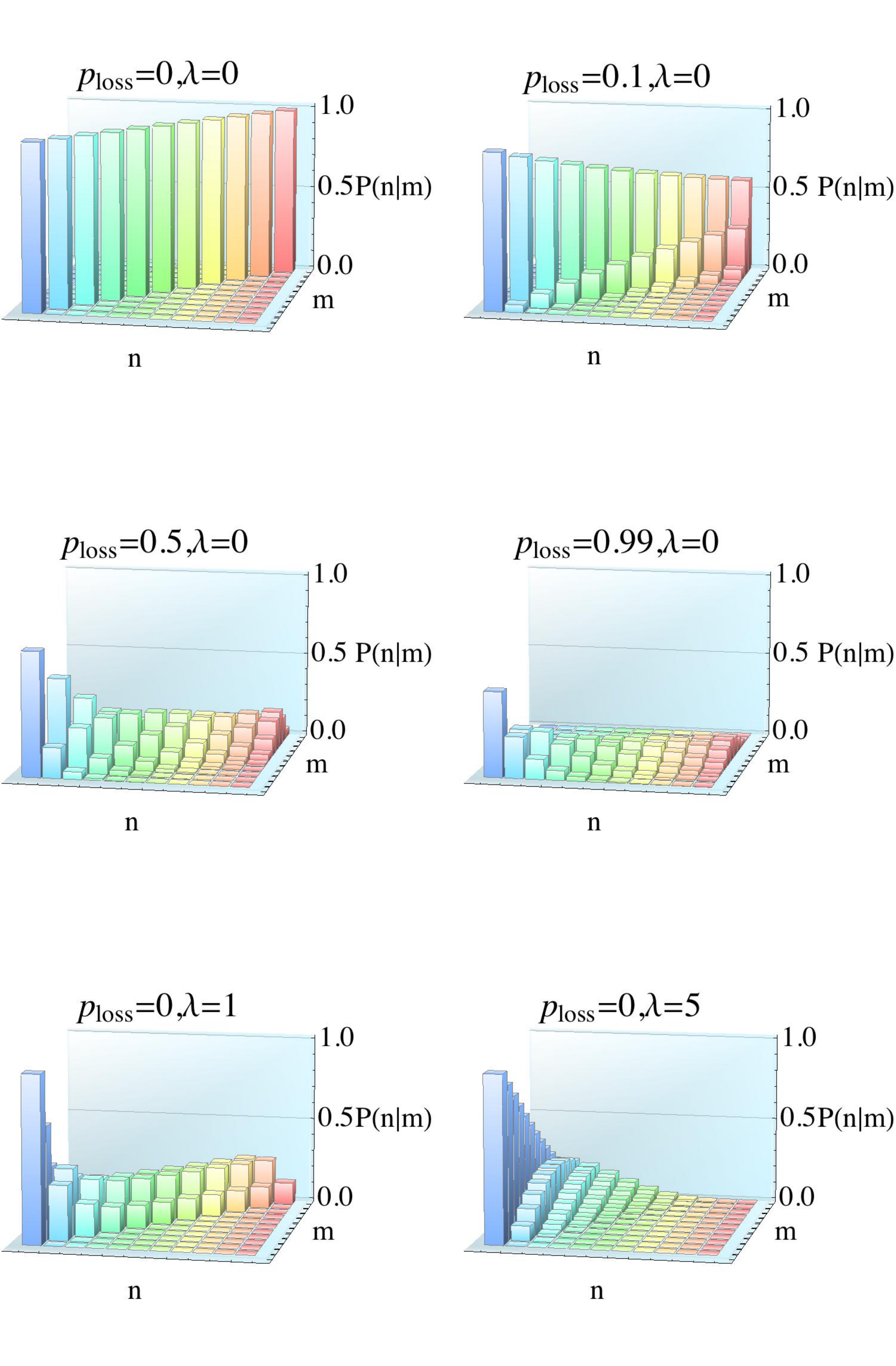}
\caption{(Colour online) $P(n|m)$ matrices for ideal, lossy and dark-county detectors, and an incident PDC state. $m=n=0$ is front-left in each inset.} \label{fig:pnm}
\end{figure}

\begin{figure}[!htb]
\includegraphics[width=\columnwidth]{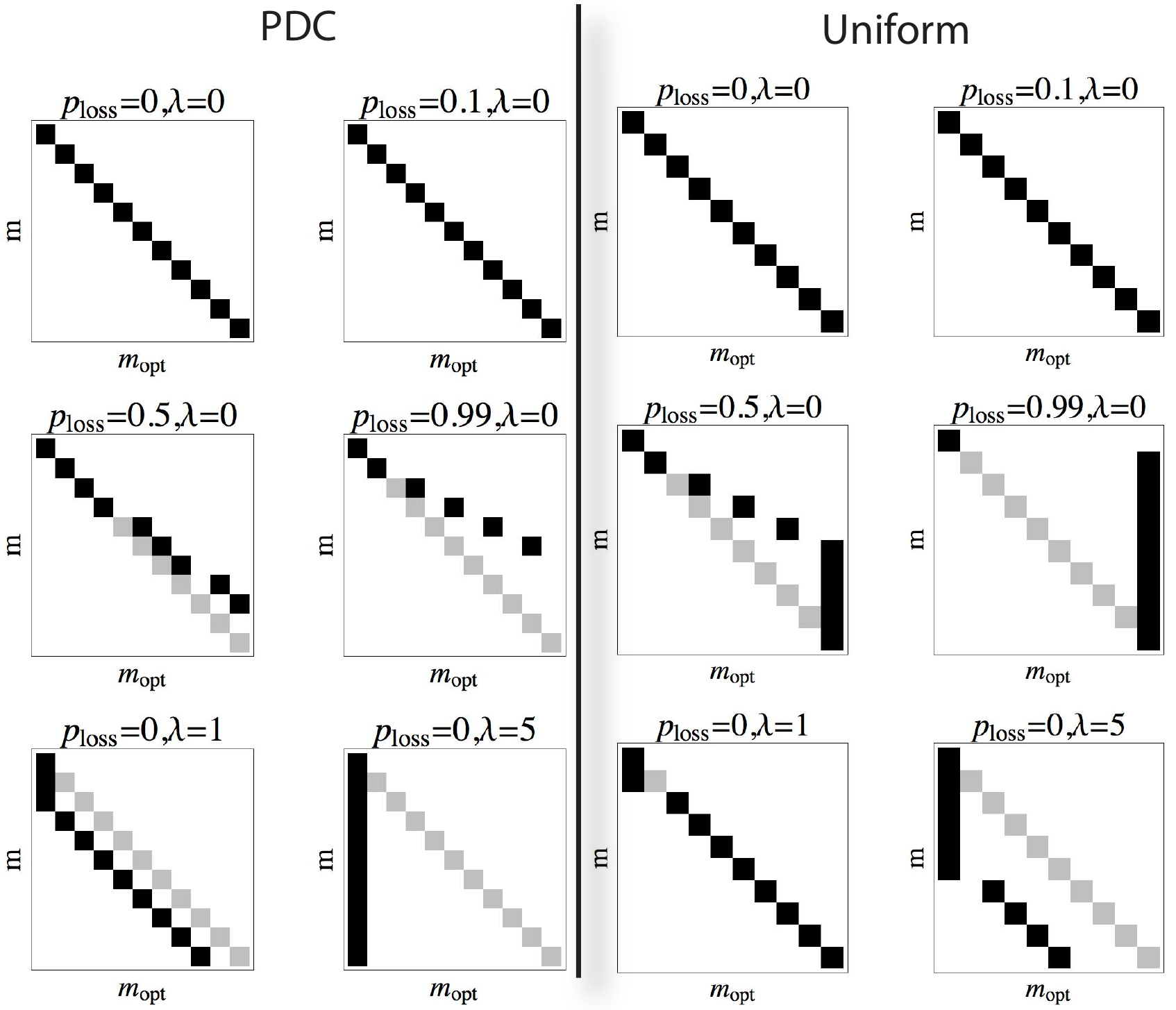}
\caption{Optimisation maps for ideal, lossy and dark-county detectors with PDC (left) and uniform (right) input states. Black boxes illustrate the mapping between raw and optimised signatures, while grey represents the identity (ideal) map and is used to guide the eye. $m=m_\mathrm{opt}=0$ is top-left in each inset.} \label{fig:maps}
\end{figure}

\begin{figure}[!htb]
\includegraphics[width=\columnwidth]{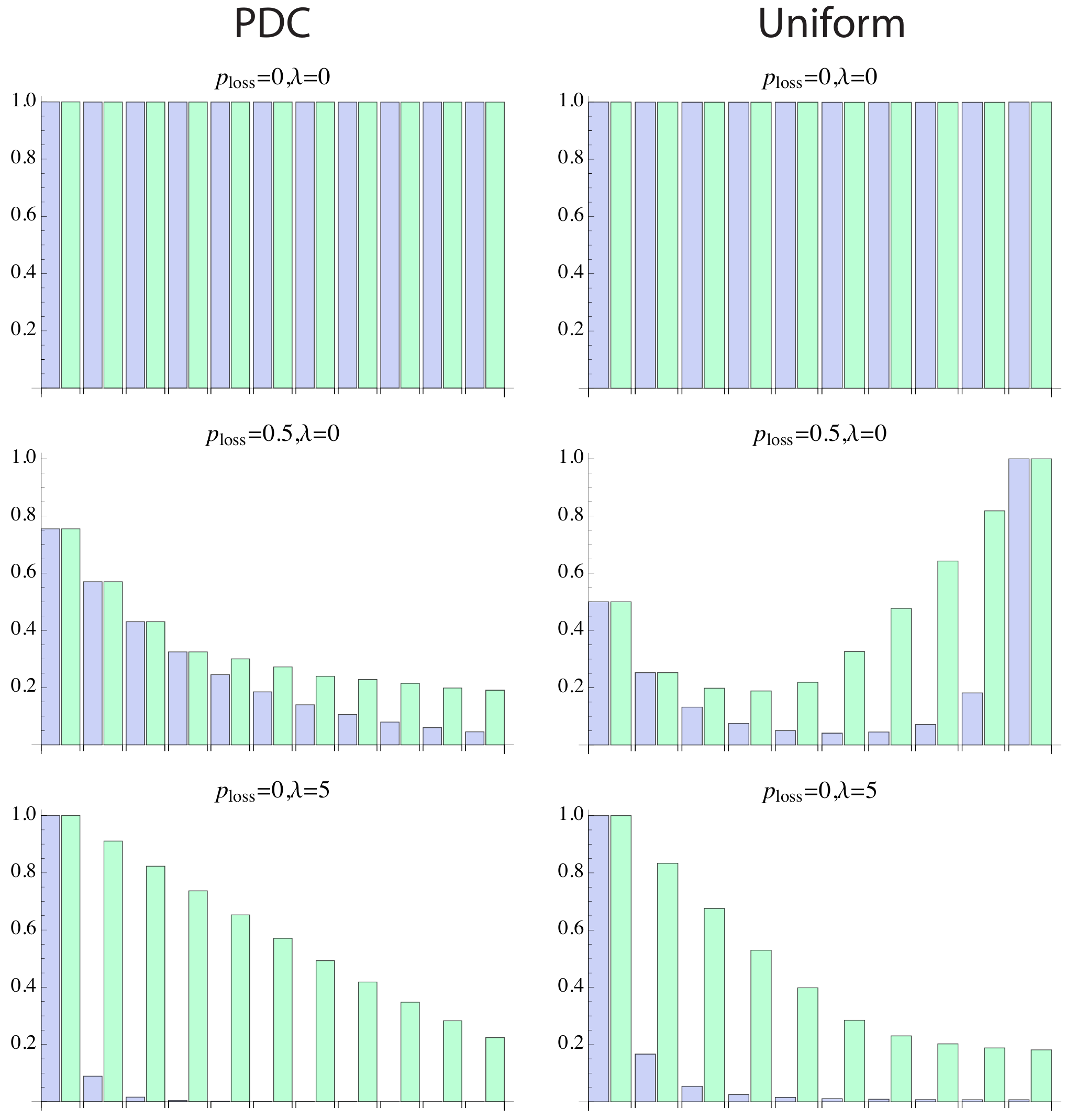}
\caption{(Colour online) Raw (blue) and optimised (green) measurement fidelities for ideal, lossy and dark-county detectors with a PDC (left) and uniform (right) input number distributions.} \label{fig:fidelity}
\end{figure}

In Fig. \ref{fig:pnm} we plot the $P(n|m)$ matrices, and in Fig. \ref{fig:maps} the corresponding optimisation maps (i.e. the mapping between the raw $m$ and $m_\mathrm{opt}$) for our detector, assuming incident PDC and uniform states. When \mbox{$p_\mathrm{loss}=p_\mathrm{DC}=0$} the detector behaves ideally and the optimisation map is the trivial identity map, i.e. the raw detector signature should be interpreted as is. As losses are introduced, $p_\mathrm{loss}>0$, it becomes desirable to compensate for the losses by incrementing the raw detector signature. On the other hand, when dark-counts are introduced, $p_\mathrm{DC}>0$, the opposite holds, and it becomes desirable to compensate for the dark-counts by decrementing the raw signature. In the limit of large dark-count rates we observe that it is optimal to always assume that $m_\mathrm{opt}=0$, since the detector is not giving us any useful information and the best we can do is guess that we have the most likely photon number, which in the case of our PDC source is $n=0$. Note that the degree of required compensation in the measurement signature is strongly dependent on the underlying photon-number distribution.

Importantly, in the regime where the optimisation map is non-trivial, the $P(n|m)$ matrix is very `washed-out'. For high fidelity measurements we desire highly localised peaks in the $P(n|m)$ matrix. However, this does not occur in the regime where optimisation may take place. Thus, the applicability of the technique is rather limited.

In Fig. \ref{fig:fidelity} we compare the difference between the fidelity of the raw and optimised detector signatures with PDC and uniform input states. In the case of \mbox{$p_\mathrm{loss}=p_\mathrm{DC}=0$} we have ideal detector operation and there is no optimisation. However, as losses and dark-counts are introduced, the optimisation technique improves the measurement fidelity. The bias in the fidelity plots are very distinct for different photon-number distributions.

\section{Conclusion}

In conclusion, we have described a simple technique which may be employed to improve the confidence of \emph{single shot} signatures generated from number-resolving photo-detectors. With sufficiently high loss or dark-count rates the technique can improve the confidence of the detection result. However, this implies that the technique only works in the regime where the detector is poor. This places limitations on when this approach might be employed. However, many present-day experiments operate in the poor detector regime and this technique may nonetheless find applicability. The example we presented was for the trivial case of an experiment consisting of a single photo-detector. However this technique can easily be generalised to multiple detectors using the same Bayesian techniques. Additionally, any detector model can be employed whereby $P(m|n)$ can be characterised, extending the applicability of this approach to many detector technologies, such as visible light photon counters (VLPCs) \cite{bib:Kim99,bib:Takeuchi99}, time-multiplexed detectors (TMDs) \cite{bib:Fitch03,bib:Haderka03,bib:Achilles03,bib:Rehacek03,bib:Achilles04} or other multiplexed detection approaches. We demonstrated that the optimisation map, which determines how detector signatures should be compensated, is highly dependent on both the detector parameters and the underlying photon-number distribution. Thus, the applicability of this technique will be highly dependent on the specific application.

\begin{acknowledgments}
We thank Timothy Ralph, Malte Avenhaus and Pieter Kok for very helpful discussions. This research was conducted by the Australian Research Council Centre of Excellence for Quantum Computation and Communication Technology (Project number CE110001027).
\end{acknowledgments}

\bibliography{bibliography}

\begin{thebibliography}{15}
\expandafter\ifx\csname natexlab\endcsname\relax\def\natexlab#1{#1}\fi
\expandafter\ifx\csname bibnamefont\endcsname\relax
  \def\bibnamefont#1{#1}\fi
\expandafter\ifx\csname bibfnamefont\endcsname\relax
  \def\bibfnamefont#1{#1}\fi
\expandafter\ifx\csname citenamefont\endcsname\relax
  \def\citenamefont#1{#1}\fi
\expandafter\ifx\csname url\endcsname\relax
  \def\url#1{\texttt{#1}}\fi
\expandafter\ifx\csname urlprefix\endcsname\relax\def\urlprefix{URL }\fi
\providecommand{\bibinfo}[2]{#2}
\providecommand{\eprint}[2][]{\url{#2}}

\bibitem[{\citenamefont{Knill et~al.}(2001)\citenamefont{Knill, Laflamme, and
  Milburn}}]{bib:KLM01}
\bibinfo{author}{\bibfnamefont{E.}~\bibnamefont{Knill}},
  \bibinfo{author}{\bibfnamefont{R.}~\bibnamefont{Laflamme}}, \bibnamefont{and}
  \bibinfo{author}{\bibfnamefont{G.}~\bibnamefont{Milburn}},
  \bibinfo{journal}{Nature (London)} \textbf{\bibinfo{volume}{409}},
  \bibinfo{pages}{46} (\bibinfo{year}{2001}).

\bibitem[{\citenamefont{Kok et~al.}(2005)\citenamefont{Kok, Munro, Nemoto,
  Ralph, Dowling, and Milburn}}]{bib:Kok05}
\bibinfo{author}{\bibfnamefont{P.}~\bibnamefont{Kok}},
  \bibinfo{author}{\bibfnamefont{W.~J.} \bibnamefont{Munro}},
  \bibinfo{author}{\bibfnamefont{K.}~\bibnamefont{Nemoto}},
  \bibinfo{author}{\bibfnamefont{T.~C.} \bibnamefont{Ralph}},
  \bibinfo{author}{\bibfnamefont{J.~P.} \bibnamefont{Dowling}},
  \bibnamefont{and} \bibinfo{author}{\bibfnamefont{G.~J.}
  \bibnamefont{Milburn}}, \bibinfo{journal}{Rev. Mod. Phys.}
  \textbf{\bibinfo{volume}{79}}, \bibinfo{pages}{135} (\bibinfo{year}{2005}).

\bibitem[{\citenamefont{Kok and Lovett}(2010)}]{bib:KokLovett11}
\bibinfo{author}{\bibfnamefont{P.}~\bibnamefont{Kok}} \bibnamefont{and}
  \bibinfo{author}{\bibfnamefont{B.~W.} \bibnamefont{Lovett}},
  \emph{\bibinfo{title}{Introduction to Optical Quantum Information
  Processing}} (\bibinfo{publisher}{Cambridge Press}, \bibinfo{year}{2010}).

\bibitem[{\citenamefont{Nielsen and Chuang}(2000)}]{bib:NielsenChuang00}
\bibinfo{author}{\bibfnamefont{M.~A.} \bibnamefont{Nielsen}} \bibnamefont{and}
  \bibinfo{author}{\bibfnamefont{I.~L.} \bibnamefont{Chuang}},
  \emph{\bibinfo{title}{Quantum Computation and Quantum Information}}
  (\bibinfo{publisher}{Cambridge University Press, Cambridge},
  \bibinfo{year}{2000}).

\bibitem[{\citenamefont{Lee et~al.}(2004)\citenamefont{Lee, Yurtsever, Kok,
  Hockney, Adami, Braunstein, and Dowling}}]{bib:Lee04}
\bibinfo{author}{\bibfnamefont{H.}~\bibnamefont{Lee}},
  \bibinfo{author}{\bibfnamefont{U.}~\bibnamefont{Yurtsever}},
  \bibinfo{author}{\bibfnamefont{P.}~\bibnamefont{Kok}},
  \bibinfo{author}{\bibfnamefont{G.~M.} \bibnamefont{Hockney}},
  \bibinfo{author}{\bibfnamefont{C.}~\bibnamefont{Adami}},
  \bibinfo{author}{\bibfnamefont{S.~L.} \bibnamefont{Braunstein}},
  \bibnamefont{and} \bibinfo{author}{\bibfnamefont{J.~P.}
  \bibnamefont{Dowling}}, \bibinfo{journal}{J. Mod. Opt.}
  \textbf{\bibinfo{volume}{51}}, \bibinfo{pages}{1517} (\bibinfo{year}{2004}).

\bibitem[{\citenamefont{Achilles et~al.}(2004)\citenamefont{Achilles,
  Silberhorn, Sliwa, Banaszek, Walmsley, Fitch, Jacobs, Pittman, and
  Franson}}]{bib:Achilles04}
\bibinfo{author}{\bibfnamefont{D.}~\bibnamefont{Achilles}},
  \bibinfo{author}{\bibfnamefont{C.}~\bibnamefont{Silberhorn}},
  \bibinfo{author}{\bibfnamefont{C.}~\bibnamefont{Sliwa}},
  \bibinfo{author}{\bibfnamefont{K.}~\bibnamefont{Banaszek}},
  \bibinfo{author}{\bibfnamefont{I.~A.} \bibnamefont{Walmsley}},
  \bibinfo{author}{\bibfnamefont{M.~J.} \bibnamefont{Fitch}},
  \bibinfo{author}{\bibfnamefont{B.~C.} \bibnamefont{Jacobs}},
  \bibinfo{author}{\bibfnamefont{T.~B.} \bibnamefont{Pittman}},
  \bibnamefont{and} \bibinfo{author}{\bibfnamefont{J.~D.}
  \bibnamefont{Franson}}, \bibinfo{journal}{J. Mod. Opt.}
  \textbf{\bibinfo{volume}{51}}, \bibinfo{pages}{1499} (\bibinfo{year}{2004}).

\bibitem[{\citenamefont{O'Sullivan et~al.}(2008)\citenamefont{O'Sullivan, Chan,
  Lakshminarayanan, and Boyd}}]{bib:OSullivan08}
\bibinfo{author}{\bibfnamefont{M.~N.} \bibnamefont{O'Sullivan}},
  \bibinfo{author}{\bibfnamefont{K.~W.~C.} \bibnamefont{Chan}},
  \bibinfo{author}{\bibfnamefont{V.}~\bibnamefont{Lakshminarayanan}},
  \bibnamefont{and} \bibinfo{author}{\bibfnamefont{R.~W.} \bibnamefont{Boyd}},
  \bibinfo{journal}{Phys. Rev. A} \textbf{\bibinfo{volume}{77}},
  \bibinfo{pages}{023804} (\bibinfo{year}{2008}).

\bibitem[{\citenamefont{Rohde and Ralph}(2006)}]{bib:RohdeRalph06b}
\bibinfo{author}{\bibfnamefont{P.~P.} \bibnamefont{Rohde}} \bibnamefont{and}
  \bibinfo{author}{\bibfnamefont{T.~C.} \bibnamefont{Ralph}},
  \bibinfo{journal}{J. Mod. Opt.} \textbf{\bibinfo{volume}{53}},
  \bibinfo{pages}{1589} (\bibinfo{year}{2006}).

\bibitem[{\citenamefont{Rohde et~al.}(2007)\citenamefont{Rohde, Webb,
  Huntington, and Ralph}}]{bib:RohdeWebb07}
\bibinfo{author}{\bibfnamefont{P.~P.} \bibnamefont{Rohde}},
  \bibinfo{author}{\bibfnamefont{J.~G.} \bibnamefont{Webb}},
  \bibinfo{author}{\bibfnamefont{E.~H.} \bibnamefont{Huntington}},
  \bibnamefont{and} \bibinfo{author}{\bibfnamefont{T.~C.} \bibnamefont{Ralph}}
  (\bibinfo{year}{2007}), \eprint{arXiv:0705.4003}.

\bibitem[{\citenamefont{Kim et~al.}(1999)\citenamefont{Kim, Takeuchi, Yamamoto,
  and Hogue}}]{bib:Kim99}
\bibinfo{author}{\bibfnamefont{J.}~\bibnamefont{Kim}},
  \bibinfo{author}{\bibfnamefont{S.}~\bibnamefont{Takeuchi}},
  \bibinfo{author}{\bibfnamefont{Y.}~\bibnamefont{Yamamoto}}, \bibnamefont{and}
  \bibinfo{author}{\bibfnamefont{H.~H.} \bibnamefont{Hogue}},
  \bibinfo{journal}{App. Phys. Lett.} \textbf{\bibinfo{volume}{74}},
  \bibinfo{pages}{902} (\bibinfo{year}{1999}).

\bibitem[{\citenamefont{Takeuchi et~al.}(1999)\citenamefont{Takeuchi, Kim,
  Yamamoto, and Hogue}}]{bib:Takeuchi99}
\bibinfo{author}{\bibfnamefont{S.}~\bibnamefont{Takeuchi}},
  \bibinfo{author}{\bibfnamefont{J.}~\bibnamefont{Kim}},
  \bibinfo{author}{\bibfnamefont{Y.}~\bibnamefont{Yamamoto}}, \bibnamefont{and}
  \bibinfo{author}{\bibfnamefont{H.~H.} \bibnamefont{Hogue}},
  \bibinfo{journal}{App. Phys. Lett.} \textbf{\bibinfo{volume}{74}},
  \bibinfo{pages}{1063} (\bibinfo{year}{1999}).

\bibitem[{\citenamefont{Fitch et~al.}(2003)\citenamefont{Fitch, Jacobs,
  Pittman, and Franson}}]{bib:Fitch03}
\bibinfo{author}{\bibfnamefont{M.~J.} \bibnamefont{Fitch}},
  \bibinfo{author}{\bibfnamefont{B.~C.} \bibnamefont{Jacobs}},
  \bibinfo{author}{\bibfnamefont{T.~B.} \bibnamefont{Pittman}},
  \bibnamefont{and} \bibinfo{author}{\bibfnamefont{J.~D.}
  \bibnamefont{Franson}}, \bibinfo{journal}{Phys. Rev. A}
  \textbf{\bibinfo{volume}{68}}, \bibinfo{pages}{043814}
  (\bibinfo{year}{2003}).

\bibitem[{\citenamefont{Haderka et~al.}(2003)\citenamefont{Haderka, Hamar, and
  Perina}}]{bib:Haderka03}
\bibinfo{author}{\bibfnamefont{O.}~\bibnamefont{Haderka}},
  \bibinfo{author}{\bibfnamefont{M.}~\bibnamefont{Hamar}}, \bibnamefont{and}
  \bibinfo{author}{\bibfnamefont{J.}~\bibnamefont{Perina}},
  \bibinfo{journal}{Euro. Phys. J. D} \textbf{\bibinfo{volume}{28}},
  \bibinfo{pages}{149} (\bibinfo{year}{2003}).

\bibitem[{\citenamefont{Achilles et~al.}(2003)\citenamefont{Achilles,
  Silberhorn, \'Sliwa, Banaszek, and Walmsley}}]{bib:Achilles03}
\bibinfo{author}{\bibfnamefont{D.}~\bibnamefont{Achilles}},
  \bibinfo{author}{\bibfnamefont{C.}~\bibnamefont{Silberhorn}},
  \bibinfo{author}{\bibfnamefont{C.}~\bibnamefont{\'Sliwa}},
  \bibinfo{author}{\bibfnamefont{K.}~\bibnamefont{Banaszek}}, \bibnamefont{and}
  \bibinfo{author}{\bibfnamefont{I.~A.} \bibnamefont{Walmsley}},
  \bibinfo{journal}{Opt. Lett.} \textbf{\bibinfo{volume}{28}},
  \bibinfo{pages}{2387} (\bibinfo{year}{2003}).

\bibitem[{\citenamefont{Rehacek et~al.}(2003)\citenamefont{Rehacek, Hradil,
  Haderka, Perina, and Hamar}}]{bib:Rehacek03}
\bibinfo{author}{\bibfnamefont{J.}~\bibnamefont{Rehacek}},
  \bibinfo{author}{\bibfnamefont{Z.}~\bibnamefont{Hradil}},
  \bibinfo{author}{\bibfnamefont{O.}~\bibnamefont{Haderka}},
  \bibinfo{author}{\bibfnamefont{J.}~\bibnamefont{Perina}}, \bibnamefont{and}
  \bibinfo{author}{\bibfnamefont{M.}~\bibnamefont{Hamar}},
  \bibinfo{journal}{Phys. Rev. A} \textbf{\bibinfo{volume}{67}},
  \bibinfo{pages}{061801} (\bibinfo{year}{2003}).

\end{thebibliography}

\end{document}